\documentclass[10pt,conference]{IEEEtran}
\usepackage{cite}
\usepackage{amsmath,amssymb,amsfonts}
\usepackage{graphicx}
\usepackage{textcomp}
\usepackage{xcolor}
\usepackage[hyphens]{url}
\usepackage{makecell,booktabs}
\usepackage[ruled,vlined]{algorithm2e}
\usepackage{pifont}

\def\BibTeX{{\rm B\kern-.05em{\sc i\kern-.025em b}\kern-.08em
    T\kern-.1667em\lower.7ex\hbox{E}\kern-.125emX}}

\title{AP-DRL: A Synergistic Algorithm-Hardware Framework for Automatic Task Partitioning of Deep Reinforcement Learning on Versal ACAP}

\author{\IEEEauthorblockN{Enlai Li\IEEEauthorrefmark{1}\IEEEauthorrefmark{4},
Zhe Lin\IEEEauthorrefmark{2},
Sharad Sinha\IEEEauthorrefmark{3}, 
and Wei Zhang\IEEEauthorrefmark{1}}
\IEEEauthorblockA{\IEEEauthorrefmark{1}
The Hong Kong University of Science and Technology, Hong Kong SAR, China}
\IEEEauthorblockA{\IEEEauthorrefmark{2}Sun Yat-Sen Univeristy, Guangdong, China}
\IEEEauthorblockA{\IEEEauthorrefmark{3}Indian Institute of Technology Goa, Goa, India}
\IEEEauthorblockA{\IEEEauthorrefmark{4}Corresponding Author (enlai.li@connect.ust.hk)}}

\begin{document}
\maketitle
\thispagestyle{plain}
\pagestyle{plain}


\begin{abstract}

Deep reinforcement learning has demonstrated remarkable success across various domains. However, the tight coupling between training and inference processes makes accelerating DRL training an essential challenge for DRL optimization. Two key issues hinder efficient DRL training: (1) the significant variation in computational intensity across different DRL algorithms and even among operations within the same algorithm complicates hardware platform selection, while (2) DRL's wide dynamic range could lead to substantial reward errors with conventional FP16+FP32 mixed-precision quantization. While existing work has primarily focused on accelerating DRL for specific computing units or optimizing inference-stage quantization, we propose AP-DRL to address the above challenges.

AP-DRL is an automatic task partitioning framework that harnesses the heterogeneous architecture of AMD Versal ACAP (integrating CPUs, FPGAs, and AI Engines) to accelerate DRL training through intelligent hardware-aware optimization. Our approach begins with bottleneck analysis of CPU, FPGA, and AIE performance across diverse DRL workloads, informing the design principles for AP-DRL's inter-component task partitioning and quantization optimization. The framework then addresses the challenge of platform selection through design space exploration-based profiling and ILP-based partitioning models that match operations to optimal computing units based on their computational characteristics. For the quantization challenge, AP-DRL employs a hardware-aware algorithm coordinating FP32 (CPU), FP16 (FPGA/DSP), and BF16 (AI Engine) operations by leveraging Versal ACAP's native support for these precision formats. Comprehensive experiments indicate that AP-DRL can achieve speedup of up to 4.17$\times$ over programmable logic and up to 3.82$\times$ over AI Engine baselines while maintaining training convergence.

\end{abstract}
\section{Introduction}

Recent years have witnessed the widespread adoption of Reinforcement Learning (RL), particularly Deep Reinforcement Learning (DRL), across diverse domains including control systems, gaming, Reinforcement Learning with Human Feedback (RLHF), and Electronic Design Automation (EDA) \cite{intro_rlhf}\cite{intro_control}\cite{intro_eda}\cite{intro_game}. Unlike supervised deep learning approaches, DRL exhibits a tight coupling between the training and inference processes, where the training data originates from both the inference and training operations themselves. This inherent characteristic makes the optimization of the training process a critical consideration for DRL deployment and optimization.


However, DRL training faces two significant challenges: \textbf{First}, selecting a universal platform for diverse DRL algorithms is non-trivial. The computational intensity varies significantly not only across different DRL algorithms but also among operations within the same algorithm. This forces DRL acceleration designs to make platform-specific tradeoffs: For low-computation-intensity scenarios (e.g., small-batch training), FPGAs are often preferred over GPUs to avoid high initialization overhead and low GPU utilization \cite{gpu_unsuitable1, gpu_unsuitable2, drl1, replay_buf1}. Conversely, for high-computation-intensity DRL workloads (e.g., CNN-based models), FPGA acceleration is frequently constrained by their low clock frequencies. Thus, identifying a universal platform for diverse DRL scenarios remains challenging. \textbf{Second}, quantization optimization for DRL training is inherently difficult. DRL applications typically exhibit a wider dynamic range distribution compared to supervised learning, making them more sensitive to data range variations \cite{quarl}. Conventional mixed-precision training based on 16-bit (FP16) and 32-bit floating-point (FP32) formats may introduce significant reward errors.


The growing adoption of heterogeneous systems for optimizing machine learning has positioned AMD's Versal Adaptive Compute Acceleration Platform (ACAP) as a promising solution to the aforementioned challenges. \textbf{First}, for the hardware selection challenge, Versal ACAP integrates multiple processing units, including CPUs, FPGA fabric, and AI Engines (AIEs), enabling DRL algorithms to dynamically leverage different compute resources based on their specific computational requirements. This architectural flexibility eliminates the need for platform trade-offs that conventional implementations face. \textbf{Second}, regarding quantization optimization, Versal ACAP's AI Engine-Machine Learning (AIE-ML) architecture provides native hardware support for Brain Float 16 (BF16) format. Crucially, BF16 maintains the same exponent range as full-precision FP32 while reducing memory and computational overhead \cite{bf16_2}\cite{bf16}. This characteristic makes it particularly suitable for DRL applications, as it preserves numerical precision while accommodating the wide dynamic ranges of DRL workloads.

However, employing Versal ACAP for DRL acceleration raises several critical research questions: (1) How to optimally map computational tasks to processing units based on both the distinctive characteristics of different DRL algorithms and the varying operational requirements within the same algorithm? (2) How to design an effective task partitioning algorithm that allocates operations to their most suitable processing units while minimizing inter-component communication overhead? (3) How to develop hardware-aware quantization algorithms that enable coordinated quantization across heterogeneous compute units, specifically for DRL training?

To address these challenges, we first conduct a quantitative analysis of performance bottlenecks across Versal ACAP's CPU, FPGA, and AIE under various DRL scenarios. Building on the obtained insights, we propose AP-DRL, an automated framework featuring layer-wise partitioning based on hardware-specific performance profiling and Integer Linear Programming (ILP)-based optimization models. AP-DRL further incorporates hardware-aware quantization algorithms tailored to different processing units. Our key contributions include:

\begin{itemize}
    \item Quantitative characterization of processing units' computational preferences using metrics including initialization ratio and clock frequency;
    \item Performance profiling for intra-component computation and inter-component communication of task partitioning nodes based on modeling tools;
    \item An ILP-based automatic partitioning model using network layers as its basic partitioning nodes, balancing parallelization efficiency and hardware-specific characteristics;
    \item A novel hardware-aware quantization algorithm to enable coordinated quantization across CPU, FPGA and AIE in Versal ACAP;
    \item Comprehensive evaluation across multiple benchmarks, demonstrating AP-DRL's ability to maintain convergence while achieving speedups up to 4.17$\times$ and 3.82$\times$ over PL and AIE baselines, respectively.
\end{itemize}

The rest of this paper is organized as follows: First, Section \ref{section:bgd} introduces the fundamentals of DRL, Versal ACAP architecture, DRL quantization techniques, and related work. Section \ref{section_bottleneck} then presents our bottleneck analysis, from which we extract key design principles and identify three associated implementation challenges. Subsequently, Section \ref{section:ap-drl} comprehensively describes the whole AP-DRL framework by detailing the solutions to the above challenges in Section \ref{section:granularity_profiling}, \ref{section:partitioning_algo}, and \ref{section_quant_algo}, respectively. Following this technical exposition, Section \ref{section:experiments} systematically evaluates the system performance through extensive experiments. Finally, Section \ref{section:conclusion} concludes the work.
\section{Background and Related Work}\label{section:bgd}

\subsection{Deep Reinforcement Learning Algorithm}

\begin{figure}[tbh]
    \centering
    \includegraphics[width=8.8cm]{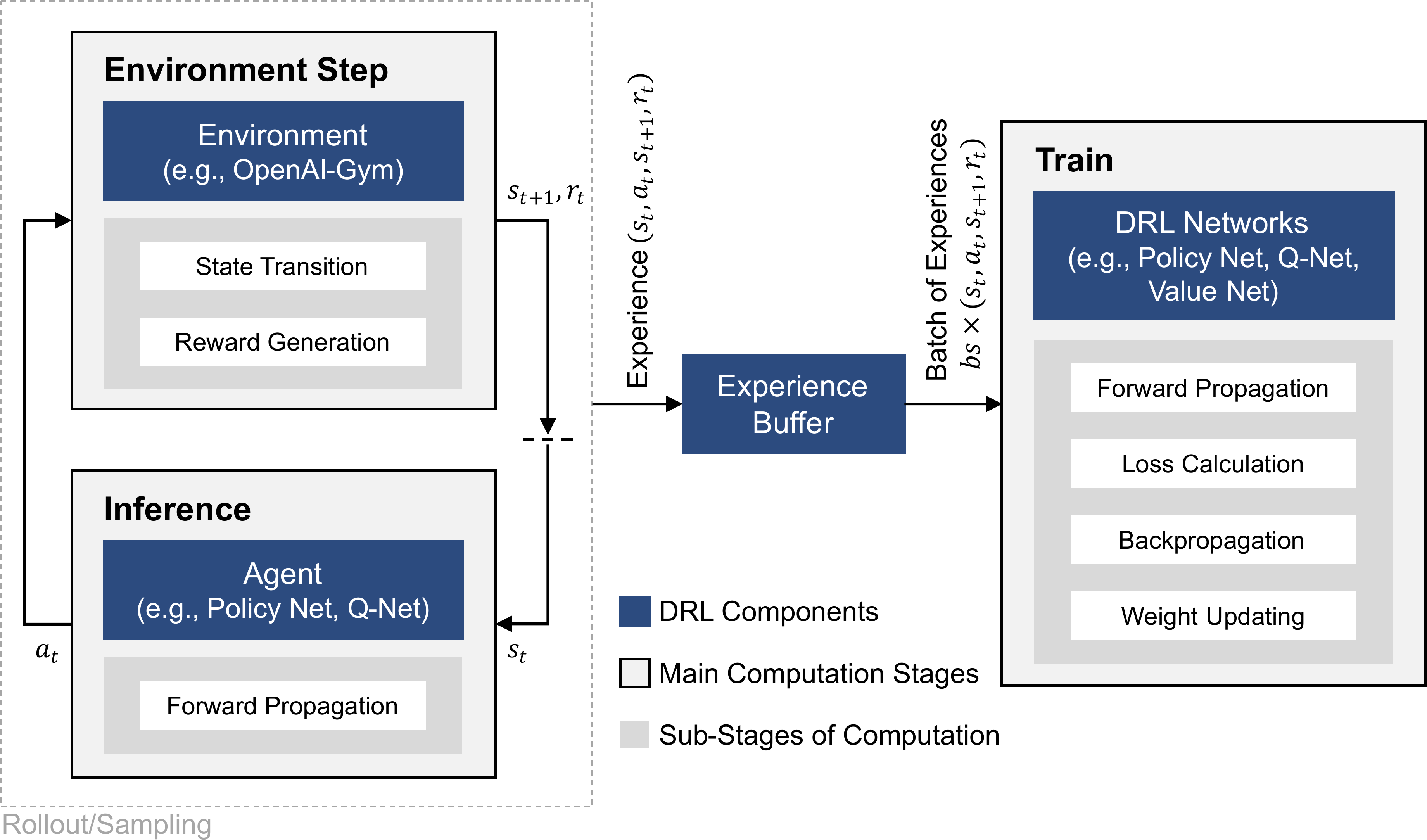}
    \caption{Workflow overview of the DRL algorithm.}
    \label{fig_drl_algo}
\end{figure}


Reinforcement Learning (RL) is traditionally formulated as a Markov Decision Process (MDP) defined by the tuple $(\mathcal{S}, \mathcal{A}, \mathcal{P}, \mathcal{R}, \gamma)$, where $\mathcal{S}$ is the state space, $\mathcal{A}$ the action space, $\mathcal{P}(s'\mid s,a)$ the transition kernel, $\mathcal{R}(s,a)$ the reward function and $\gamma \in [0,1]$ the discount factor. At each time step $t$, the agent observes a state $s_t \in \mathcal{S}$, samples an action $a_t \in \mathcal{A}$ according to policy $\pi(a \mid s)$, receives reward $r$ and the next state $s_{t+1} \sim \mathcal{P}(\cdot \mid s, a)$, and repeats the process until the termination at the time step $T$. The objective of RL is to learn a policy $\pi$ that maximizes the expected discounted return $\mathbb{E} \left[ \sum_{k=0}^{T-t} \gamma^{k} r_{t+k} \right]$.

While classic RL methods struggle with high-dimensional state and action spaces, Deep Reinforcement Learning (DRL) addresses this challenge by employing deep neural networks to approximate core components, such as value functions ($Q^\pi$), policies ($\pi$), or combined actor-critic models. Figure \ref{fig_drl_algo} illustrates the workflow overview of DRL as a repeating sequence of three main computation stages: \textbf{Inference}, \textbf{Environment Step}, and \textbf{Train}. The first stage, \textbf{Inference}, is executed by the \textit{Agent}: given the current state~$s_t$, the agent performs forward propagation to yield an action~$a_t$ according to policy $\pi(a\mid s_t)$. Immediately afterwards, \textbf{Environment Step} is triggered: the environment receives~$a_t$, performs a state transition~$s_{t+1}\sim\mathcal{P}(\cdot\mid s_t,a_t)$ and generates the corresponding reward~$r_t=\mathcal{R}(s_t,a_t)$.  
The resulting experience tuple~$(s_t,a_t,r_t,s_{t+1})$ is stored in the \textit{Experience Buffer}.  

These two stages are repeated until the buffer contains enough samples to initiate the \textbf{Train} stage, a stage that dominates the overall workflow execution time \cite{drl2}. During training, a batch with size~$bs$ is sampled from the buffer and fed into the DRL networks (e.g., Q-Net, Policy-Net, Value-Net). The networks first perform \textit{forward propagation} to produce predictions such as Q-values or action probabilities; a \textit{loss function} is then evaluated to quantify the discrepancy between predictions and targets derived from~$r_t$ and~$s_{t+1}$. Finally, \textit{backpropagation} computes the gradients, and an optimizer performs \textit{weight updates} and completes the timestep $t$. One episode of the workflow ends when $t$ equals the termination timestep $T$.

The discussion above reveals that, unlike supervised learning, DRL tightly couples inference and training phases, requiring co-optimization of both processes. For non-distributed DRL algorithms where training dominates computation time, optimizing this phase becomes particularly critical. Existing research has pursued DRL acceleration through two primary approaches. One line of work has developed specialized heterogeneous architectures combining CPUs with accelerators (GPUs/FPGAs) to accelerate the DRL inference and training pipeline \cite{drl1,drl2,drl4,cpu-gpu-fpga,fixar}. Another direction has focused on optimizing individual DRL components, including custom-designed off-chip replay buffers \cite{cpu-gpu-fpga,replay_buf1}, accelerated environment simulation \cite{env_sim}, and hardware-optimized Generalized Advantage Estimation (GAE) computation for Proximal Policy Optimization (PPO) algorithms \cite{gae}. In contrast to these previous manual design approaches, our work introduces a novel methodology combining automatic task partitioning across heterogeneous components in the AMD Versal ACAP platform with hardware-aware quantization for systematic DRL training optimization.


\subsection{The Versal ACAP Architecture}\label{section_versal_arch}

\begin{figure}
    \centering
    \includegraphics[width=8.8cm]{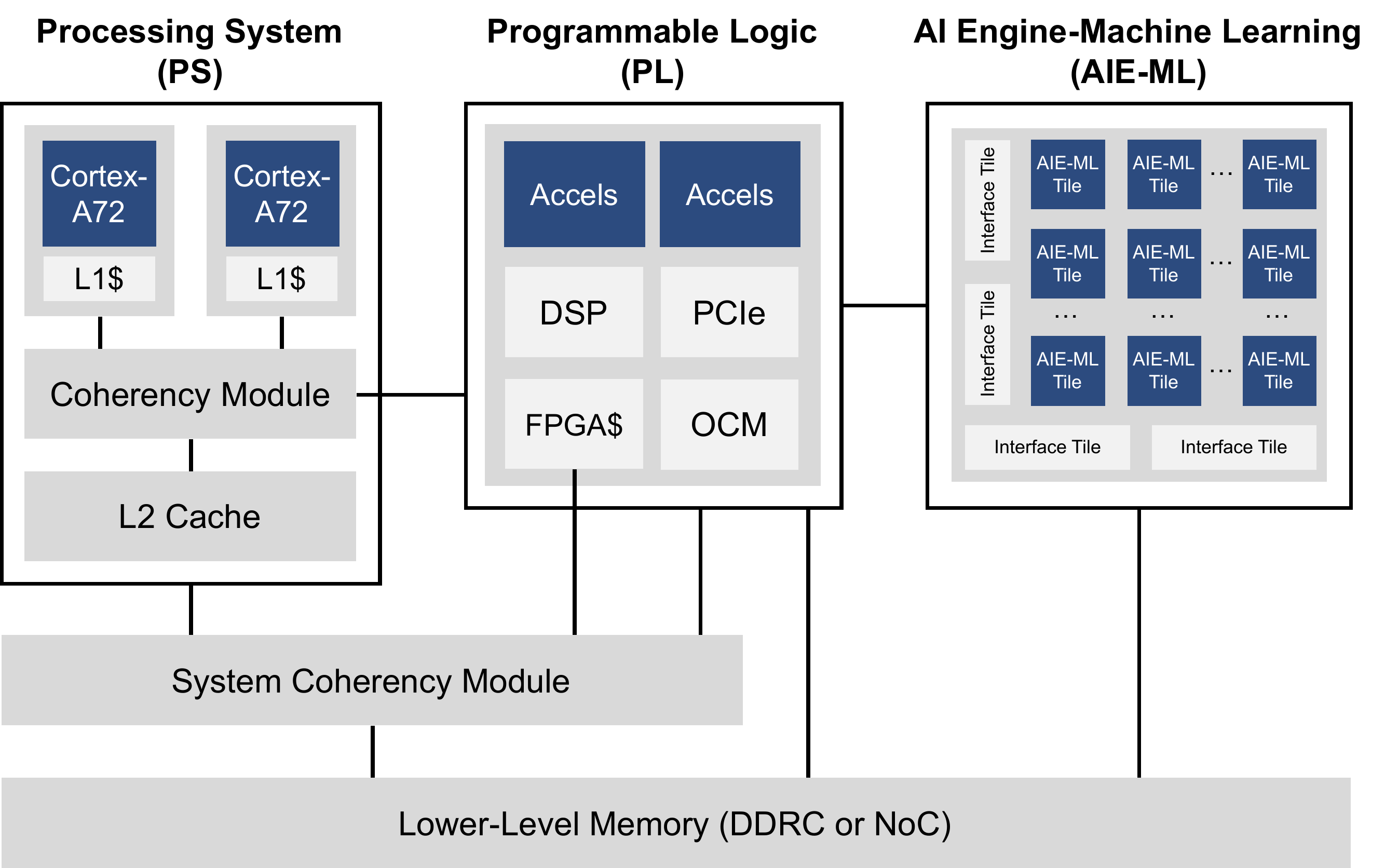}
    \caption{The overview of Versal ACAP (AI Edge series) architecture.}
    \label{fig:versal_arch}
\end{figure}

Figure \ref{fig:versal_arch} illustrates the heterogeneous architecture of the Versal ACAP (AI Edge series). The Versal ACAP primarily consists of three key components: the Processing System (PS, acts as CPUs), composed of dual-core Cortex-A72 processors and their cache hierarchy; the Programmable Logic (PL, acts as FPGA), incorporating reconfigurable units, DSP engines, and on-chip memory (OCM); and the AI Engine-Machine Learning (AIE-ML) module, which comprises an array of AIE-ML tiles and interface tiles.

The PS and PL are interconnected through various coherent shared memory architectures, facilitated by multiple 128-bit AXI interfaces. These interfaces enable the PL to access the PS's L1 cache, last-level cache, or establish a full coherency architecture by implementing the PL cache to communicate with the PS, thereby providing versatile collaboration mechanisms between the PS and PL \cite{tapca}\cite{fca}. In contrast, communication between the PL and AIE-ML is relatively streamlined, primarily conducted via Programmable Logic Input/Outputs (PLIOs) embedded within multiple interface tiles. Each interface logic includes several PL input/output interfaces operating at the PL clock frequency, alongside input/output interfaces operating at the 1 GHz AIE-ML clock frequency. This design ensures high-throughput data transfer between the PL and AIE-ML, collectively enabling efficient heterogeneous coordination across the PS, PL, and AIE-ML domains.

Compared with the AI Engine (AIE) integrated in the Versal ACAP AI Core series, the AIE-ML in the AI Edge series introduces several architectural optimizations for machine learning applications. A critical enhancement for DRL acceleration is its native support for the Brain Float 16 (BF16) data format (detailed discussions on BF16 are provided in Sections \ref{section_quant_bgd} and \ref{section_data_prec}). Due to its identical exponent precision to FP32, BF16 effectively preserves the dynamic range of weights during training, making it particularly suitable for DRL applications that are sensitive to numerical precision \cite{wider_dynamic3}\cite{wider_dynamic2}\cite{wider_dynamic1}. Moreover, since the PL/DSP provides native hardware support for both FP16 and FP32 data formats while the PS operates on FP32, the Versal ACAP can leverage efficient conversion between FP16, FP32 and BF16 to enable heterogeneous mixed-precision training optimization across PS, PL and AIE-ML components. For clarity throughout this paper, both terminologies of ``AIE" and ``AIE-ML" refer specifically to AIE-ML unless otherwise noted.

\subsection{Quantization in DRL}\label{section_quant_bgd}

Quantization, as a prevalent optimization technique in deep learning, operates by reducing data precision to decrease model memory footprints while exploiting hardware-specific precision support to enhance computational efficiency during both training and inference phases. Within supervised learning paradigms, quantization techniques are predominantly applied to accelerate inference processes \cite{qat}. However, their implementation in unsupervised DRL presents unique challenges. The intrinsic coupling of inference and training phases in DRL systems, combined with the predominant computational overhead of training in non-distributed implementations, renders conventional quantization approaches, like Post-Training Quantization (PTQ) and Quantization-Aware Training (QAT), potentially suboptimal for DRL optimization.

Mixed-precision training optimizes model performance during training by employing lower-precision data formats (e.g., FP16) for operations that are insensitive to data range, while maintaining full precision (FP32) for range-sensitive operations \cite{mixed_prec_training}. However, in DRL applications, the adoption of quantization techniques, particularly mixed-precision training, remains limited due to two key characteristics: (1) the substantially larger dynamic range of weight distributions compared to supervised deep learning models, and (2) the potential propagation of quantization errors across subsequent states during temporal processing. \cite{quarl} and \cite{actorq} \cite{quant_overview} have conducted comprehensive evaluations across various DRL algorithms and environments. Specifically, \cite{actorq} investigated the impact of PTQ and QAT on inference latency in distributed DRL systems, where inference constitutes the dominant computational overhead. Meanwhile, \cite{quarl} analyzed quantization effects from a weight distribution perspective and provided preliminary explorations of mixed-precision inference using mixed FP16/FP32 operations. Additionally, \cite{fixar} has proposed optimizing the DRL training phase by selectively using 16-bit fixed-point representations for activations within certain timesteps.

\begin{figure}
    \centering
    \includegraphics[width=8.8cm]{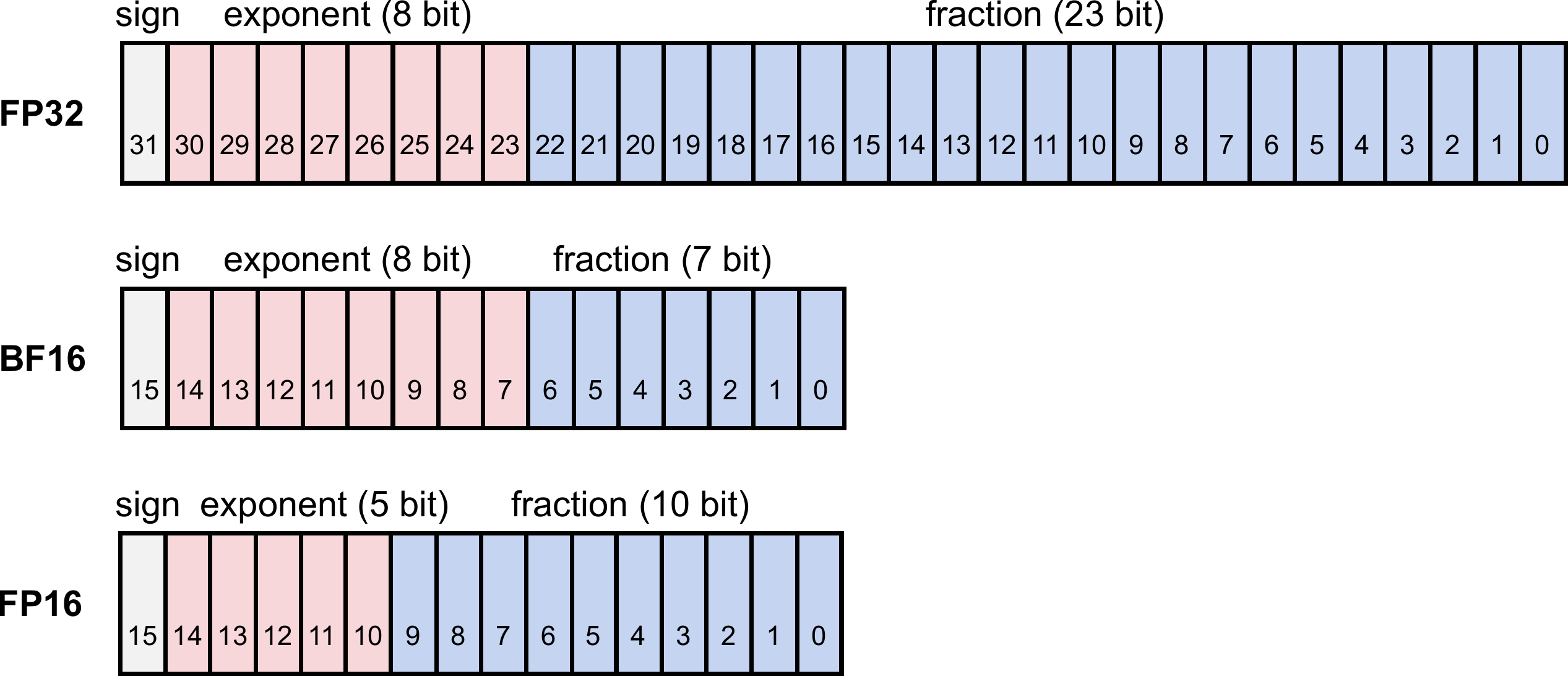}
    \caption{The binary format of BF16, FP16 and FP32.}
    \label{fig_fp32}
\end{figure}


Figure \ref{fig_fp32} illustrates the binary format of BF16, IEEE 754 half-precision floating-point format (FP16), and IEEE 754 single-precision floating-point format (FP32). Notably, unlike FP16, BF16 adopts the same exponent bits as FP32 while reducing the fraction bits. This design enables BF16 to maintain an identical data range to FP32, albeit with lower precision, making it particularly suitable for DRL applications where data range sensitivity is critical \cite{bf16_2}\cite{bf16}\cite{quarl}.

Our work proposes a hardware-aware quantization algorithm that combines master weight backup and dynamic loss scaling to enable collaborative mixed-precision optimization across the Versal ACAP’s processing units: FP16 quantization and dynamic loss scaling on the PL, BF16 quantization on the AIE, and full-precision FP32 operations on the PS. By leveraging master weight synchronization between PS-PL and PL-AIE, our design facilitates coordinated FP32+FP16+BF16 quantization, optimizing performance while maintaining DRL training stability.

\section{Bottleneck Analysis}\label{section_bottleneck}

\begin{figure}
    \centering
    \includegraphics[width=8.8cm]{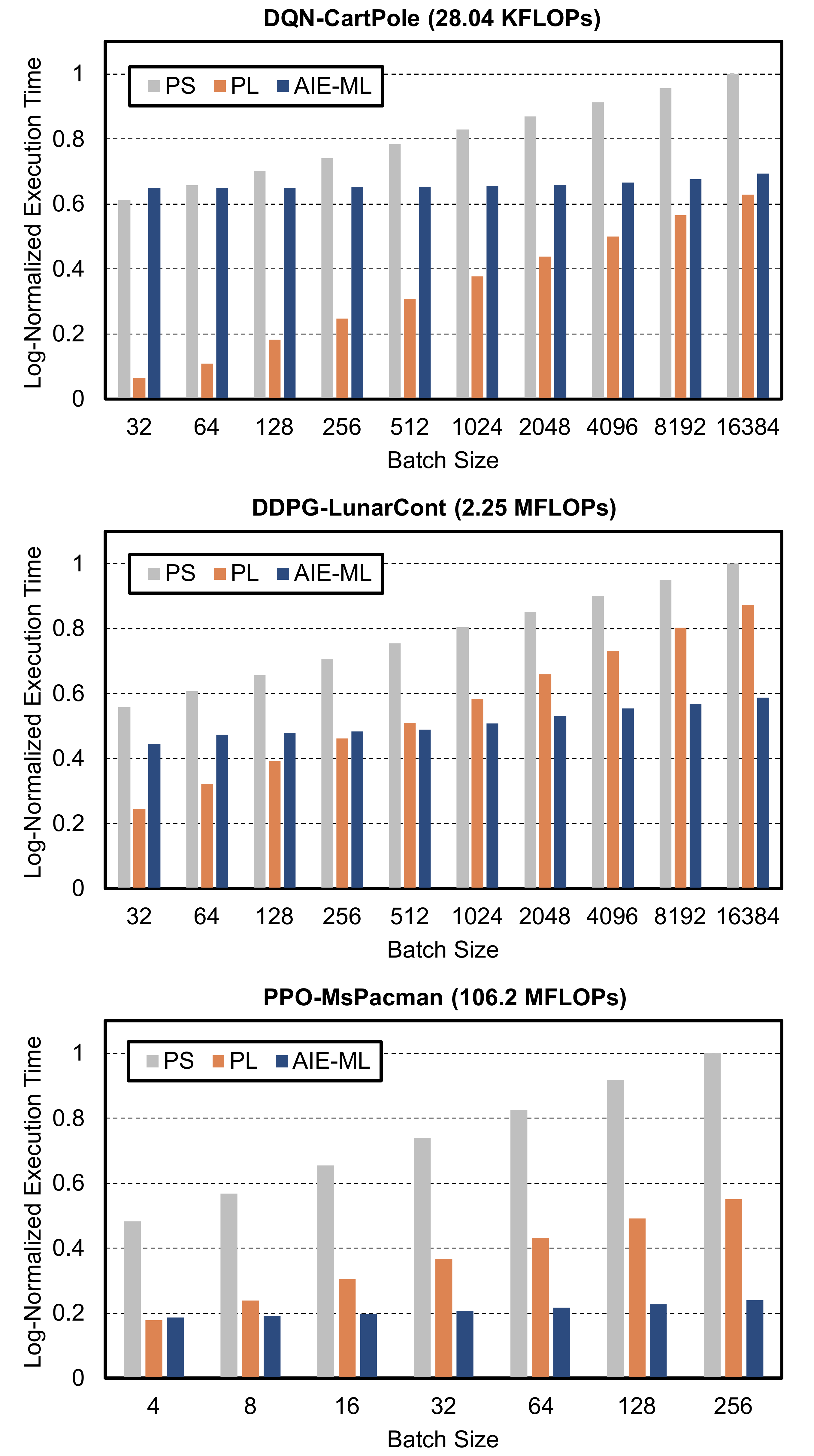}
    \caption{Log-normalized execution time of three DRL algorithm-environment combinations for single training timestep on PS, PL, and AIE-ML processors on the VEK280 platform (hardware emulation), with performance profiled by the Vitis Analyzer. Detailed configurations of these three combinations can be found in Table \ref{table_benchmark_config}.}
    \label{fig:bottleneck_psplaie}
\end{figure}

\begin{figure}
    \centering
    \includegraphics[width=8.8cm]{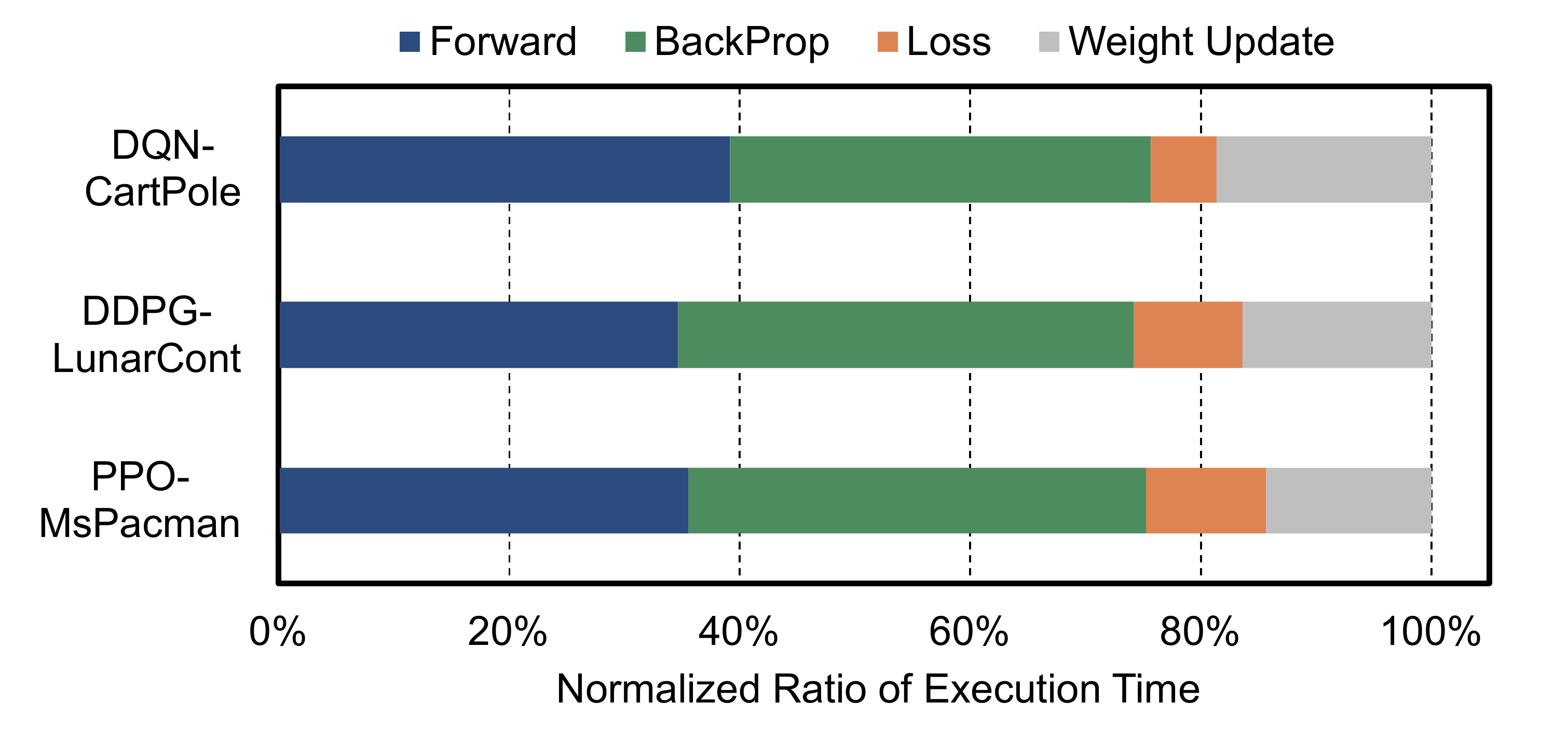}
    \caption{Execution time breakdown of one single training timestep on VEK280 PS (hardware emulation), with performance profiled through the \texttt{xiltimer} APIs. The configuration details of the three DRL algorithm-environment combinations can be found in Table \ref{table_benchmark_config}.}
    \label{fig:bottleneck_train}
\end{figure}

\begin{figure}
    \centering
    \includegraphics[width=8.8cm]{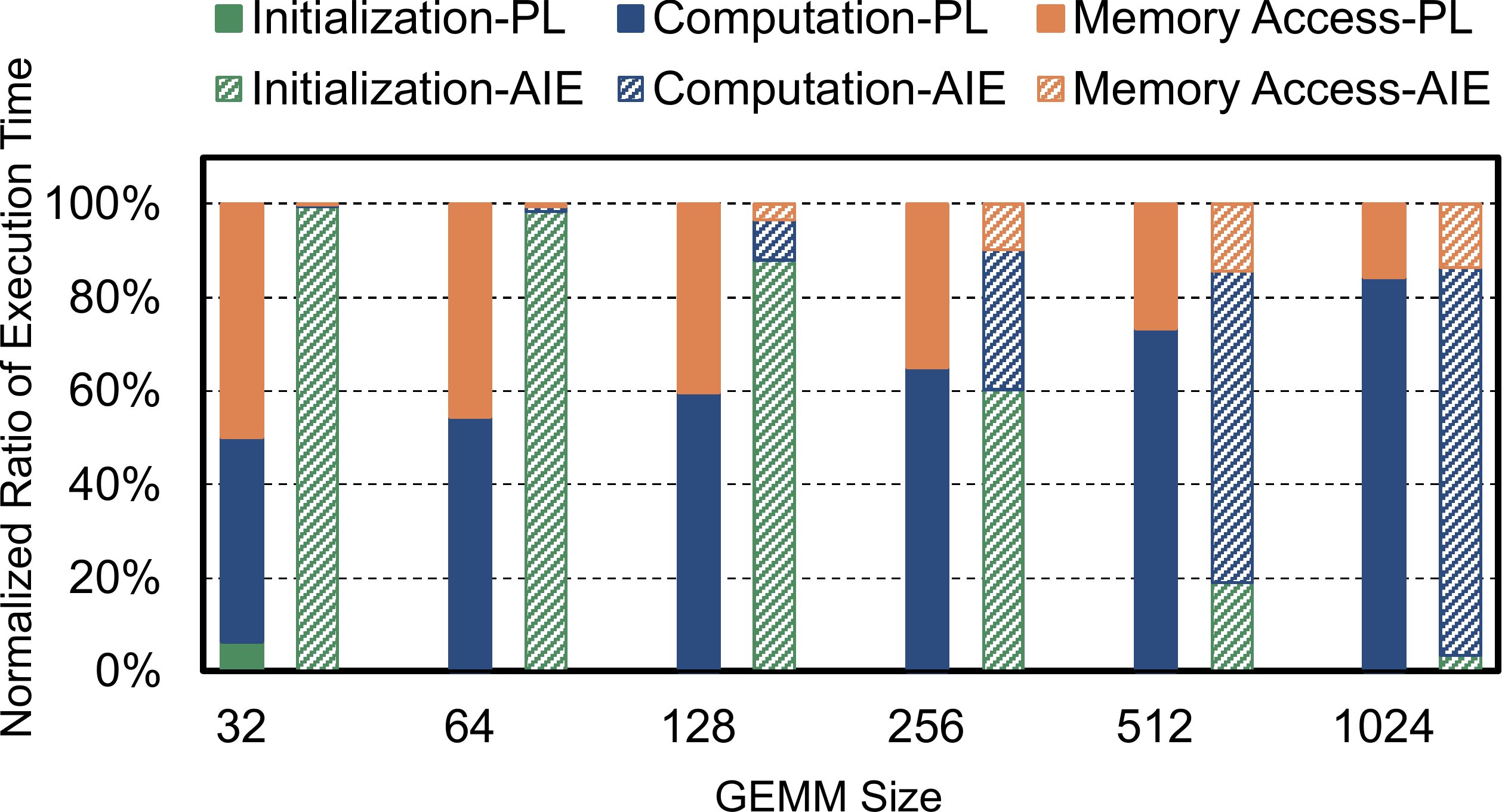}
    \caption{Execution time breakdown of the synthetic GEMMs on VEK280 PL and AIE-ML (hardware emulation), with performance profiled by the Vitis Analyzer. The clock frequencies of PL and AIE are 245 MHz and 1 GHz, respectively.}
    \label{fig:bottleneck_gemm}
\end{figure}

In this section, we first evaluate the execution time of the DRL training stage across PS, PL, and AIE under various configurations. Based on the obtained results, we then propose our systematic design principles.

\subsection{Performance Analysis of PS, PL, and AIE}

To comprehensively assess the impact of computation intensity on platform performance, we selected three DRL algorithms with distinct computation intensity and compared their training performance under different batch sizes (configuration details can be found in Table \ref{table_benchmark_config}). Computation intensity is defined as the ratio of arithmetic operations (FLOPs) to memory accesses. In the context of General Matrix Multiplication (GEMM) operations, computation intensity exhibits a positive correlation with FLOPs \cite{compute_intensity}, so we employ FLOPs as a practical substitute for computation density to simplify both theoretical analysis and experimental procedures. The environmental configurations and network architectures of the DRL algorithms are detailed in Table \ref{table_benchmark_config}. To ensure optimal utilization of each component's computational capabilities, we employed the latest performance evaluation frameworks (i.e., COMBA \cite{comba} for the PL component and CHARM \cite{charm} for the AIE-ML component) to conduct a design space exploration (DSE) of the training process, thereby guaranteeing the accuracy of experimental results. The experiments are conducted using Vitis 2023.2 on the VEK280 hardware emulation platform, with performance profiling performed by the Vitis Analyzer.

As illustrated in Figure \ref{fig:bottleneck_psplaie}, the performance across components varies significantly under different FLOPs scenarios: \ding{182} Low FLOPs scenarios: FPGAs exhibit superior performance compared to PS and AIE. \ding{183} High FLOPs scenarios: AIE-ML outperforms both PS and PL components.

To further investigate the underlying causes of these performance differences, we decompose the DRL training process within one single timestep into distinct computation phases and analyze the relative execution time of each phase, as shown in Figure \ref{fig:bottleneck_train}. The experiments are conducted via hardware emulation on the PS of the VEK280 platform, with performance profiled by the \texttt{xiltimer} APIs. The results indicate that forward propagation and backpropagation dominate the training time across all three DRL algorithms. Since both phases primarily rely on GEMM, we further examine six GEMM configurations of varying sizes (where GEMM with size $n$ denotes the multiplication of two $n\times n$ matrices) on VEK280 PL and AIE-ML through hardware emulation, with performance profiled by the Vitis Analyzer. After optimization with COMBA and CHARM, these GEMM operations are deployed on PL and AIE-ML, respectively, and their execution profiles are presented in Figure \ref{fig:bottleneck_gemm}.

Figure \ref{fig:bottleneck_gemm} reveals two main factors affecting the performance difference between PL and AIE at different computation workloads: \ding{182} For low-FLOPs operations, the AIE component's performance is limited by its relatively long kernel launch time, which accounts for a significant portion of the total execution time. In contrast, the PL component exhibits better performance due to its shorter initialization time. \ding{183} For high-FLOPs operations, the PL component's performance becomes constrained by its lower clock frequency. When examining the optimized implementations of both components while excluding initialization time, we observe a similar ratio of execution time between computation and memory access operations. However, despite this similar ratio, the PL component demonstrates inferior performance compared to AIE due to its lower clock frequency (PL@245 MHz vs. AIE@1 GHz) when handling comparable computational workloads. These observations raise an important optimization challenge for DRL applications: \textit{how to effectively leverage the complementary performance characteristics of PL and AIE components under varying computation workloads to accelerate the DRL training process?}

\subsection{Design Principles}\label{section:design_principles}

We propose the AP-DRL framework to address the above challenge. The key idea of AP-DRL involves an automated task partitioning algorithm that dynamically allocates DRL training operations to their optimal computing components, while carefully managing the inter-component communication overhead. However, this introduces several critical considerations: \textbf{First}, it is a fundamental requirement for the partitioning algorithm to determine the appropriate task granularity and obtain accurate performance profiles. \textbf{Second}, the algorithm must intelligently balance the trade-offs between parallelization efficiency and component-specific performance characteristics, which are two factors simultaneously affected by the batch size. \textbf{Third}, the framework should incorporate hardware-aware optimizations when mapping computational tasks to their target components. We will systematically address each of these challenges in the following sections detailing AP-DRL implementation.

\section{The AP-DRL Framework}\label{section:ap-drl}

In this section, we systematically address the challenges outlined in the design principles in Section \ref{section:design_principles}: we first introduce the overview of AP-DRL, and then we elaborate on the selection of partitioning granularity and the profiling process for partitioning nodes in Section \ref{section:granularity_profiling} to address the \textbf{First} challenge. \textbf{Second}, we present a partitioning model based on Integer Linear Programming (ILP) in Section \ref{section:partitioning_algo}. \textbf{Third}, the hardware-aware quantization approach is discussed in detail in Section \ref{section_quant_algo}.

\subsection{Workflow Overview of AP-DRL}\label{section:ap-drl_framework}

\begin{figure}
    \centering
    \includegraphics[width=4.5cm]{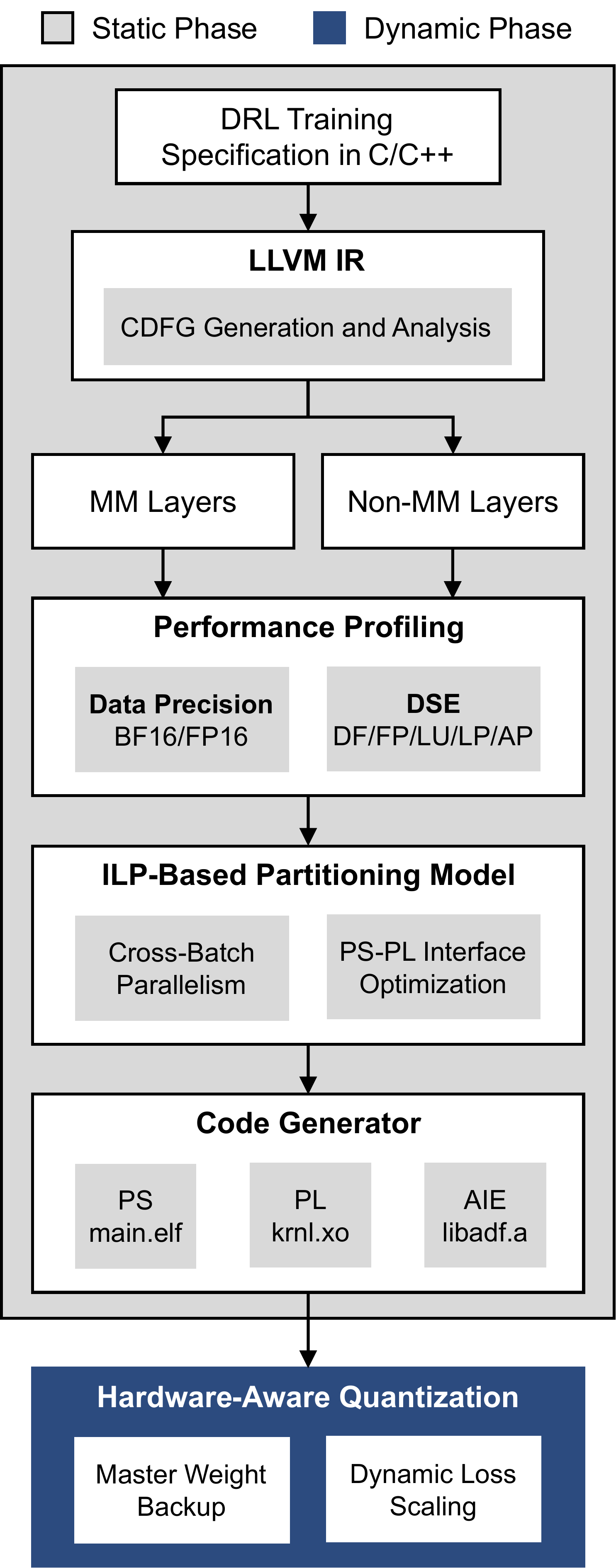}
    \caption{The overview of the AP-DRL framework.}
    \label{fig:ap-drl_framework}
\end{figure}

Figure~\ref{fig:ap-drl_framework} presents the AP-DRL framework overview. The workflow is divided into two distinct phases: a static phase and a dynamic phase. During the static phase, which executes before DRL application deployment, the system determines the optimal task partitioning strategy. The dynamic phase then performs runtime optimizations within the hardware-aware quantization based on the predetermined partitioning scheme while the DRL applications are executing.

The framework takes C/C++ DRL training process specifications as input, which are first converted to LLVM Intermediate Representation (LLVM IR) using the Clang frontend. A Control Data Flow Graph (CDFG) is then generated through LLVM Pass analysis, where layers serve as nodes with their control and data dependencies explicitly captured. This representation provides essential input for the subsequent ILP-based partitioning model.

The framework classifies layer nodes into two categories: Matrix Multiplication (MM) layers and Non-MM layers (e.g., ReLU, tanh activation functions). Non-MM layers, being unsuitable for AIE acceleration due to their computational characteristics, are typically allocated to the PL \cite{charm}\cite{eq-vit}. These layers undergo FP16 format conversion followed by DSE detailed in Section \ref{section:granularity_profiling}, to determine their optimal PL configurations. For MM layers, the partitioning decision between PL and AIE is based on their characteristics described in Section \ref{section_bottleneck}. Each MM layer is quantized into both BF16 and FP16 formats, with detailed profiling performed to assess their respective resource utilization and performance on both computing components. AIE profiling precedes PL profiling since AIE operations require partial PL resources for high-speed data transfer. The collected profiling data, including dependency relationships, execution information, and resource utilization, feeds into the ILP-based partitioning model for optimal task allocation. Concurrently, TAPCA \cite{tapca} determines the optimal shared memory configuration for PS-PL communication, particularly for the Inference$\rightarrow$Experience Buffer$\rightarrow$Sampled Training Data$\rightarrow$Updated Model (on-policy DRL) pipeline, as shown in Figure~\ref{fig:quant_hw}.

The code generation phase produces the design graph \texttt{libadf.a} for AIE and hardware kernels \texttt{kernel.xo} for PL based on the resource allocation results. Following successful compilation of both AIE kernels and graphs, and PL HLS kernels, the Vitis compiler links them with the target platform to generate the \texttt{.xclbin} and \texttt{.xsa} platform description file. Finally, the host CPU executable (\texttt{main.elf}) is generated, and the complete package is assembled into a Programmable Device Image (PDI) for subsequent hardware emulation or on-board execution.

\begin{figure}
    \centering
    \includegraphics[width=8.8cm]{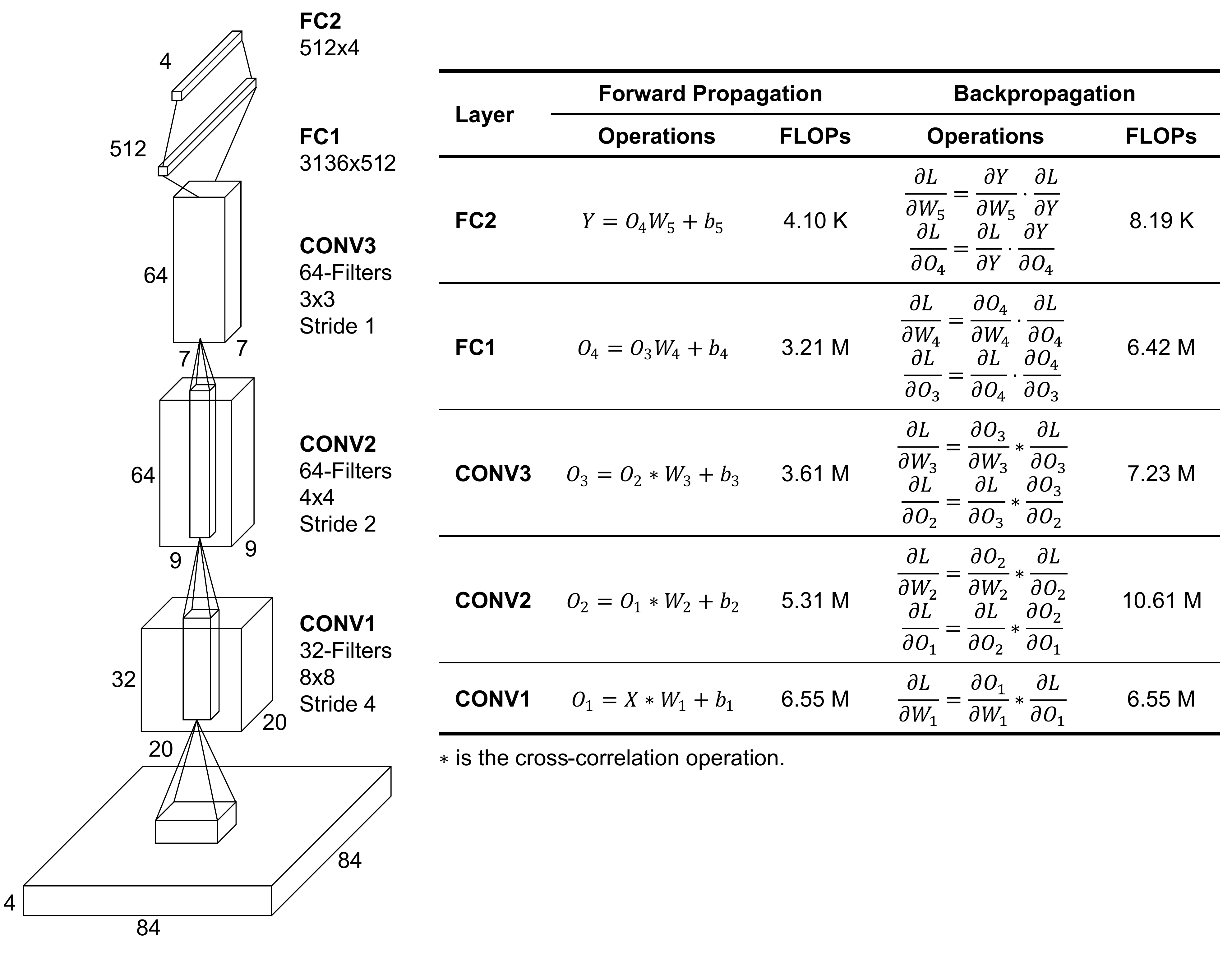}
    \caption{The network architecture and FLOPs of forward propagation and backpropagation in the DQN-Breakout environment.}
    \label{fig:methodology_layer}
\end{figure}

\subsection{Partitioning Granularity and Profiling}\label{section:granularity_profiling}

To determine the appropriate partitioning granularity, we conducted a detailed analysis of the DRL algorithm's training process and network architecture. Taking the DQN-Breakout environment as an example (detailed configurations are provided in Table \ref{table_benchmark_config}), the DQN loss function $L(\theta)$ in Equation \ref{equ_dqn_loss} reveals that computing the loss requires two forward propagation through the Q-network to obtain the target value $y$ and the online Q-value. Based on the bottleneck analysis in Section \ref{section_bottleneck}, the primary computational bottleneck in DQN-Breakout training lies in these two forward passes and one backward pass. This training pattern with multiple forward propagation and backpropagation is prevalent in DRL because decoupling target networks (as in DQN) or separating policy and value functions (as in Actor-Critic) helps stabilize training by mitigating bootstrapping bias and convergence issues.

\begin{equation}\label{equ_dqn_loss}
    L(\theta) = \mathbb{E}\left[ 
        \left( 
            \underbrace{r + \gamma \max_{a'} Q_{\text{target}}(s', a'; \theta^-)}_{\text{Target } y} 
            - \underbrace{Q(s, a; \theta)}_{\text{Online Q-value}} 
        \right)^2 
    \right]
\end{equation}

Figure \ref{fig:methodology_layer} illustrates the network architecture and the FLOPs required for a single forward and backward pass in DQN-Breakout. The training process involves computations across 15 distinct layers with FLOPs ranging from 4.10 KFLOPs to 10.61 MFLOPs, indicating significant variation in computational load per layer. This justifies layer-level partitioning from a perspective of FLOPs. Moreover, since individual layers exhibit tight intra-layer dependencies, splitting a single layer across components could introduce substantial communication overhead due to frequent inter-component data exchange. Thus, adopting a per-layer partitioning granularity in AP-DRL is a reasonable design choice.

\begin{table}[tb]
    \caption{Design Points for PU Candidates in TAPCA}
    \label{tab_design_points}
    \centering
    \begin{tabular}{llc}
      \toprule
      \textbf{Pragma} & \textbf{Configurations} & \textbf{\#Design Points}\\
      \midrule
      Dataflow (DF) & Enable/Disable & 2\\
      \midrule
      Function Pipeline (FP) & Enable/Disable & 2\\
      \midrule
      Loop Pipeline (LP) & Enable/Disable & 2\\
      \midrule
      Loop Unroll (LU) & Factors & $\lceil\log_2LB\rceil$\\
      \midrule
      Array Partition (AP) & Types \& Factors & $B_M/B_D+1$\\
      \bottomrule
    \end{tabular}
  \end{table}

To generate performance profiling data for the following partitioning algorithm, we leverage TAPCA \cite{tapca} and CHARM \cite{charm} for profiling. TAPCA and CHARM are DSE-based profiling tools for PL and AIE applications, respectively. Besides basic profiling, TAPCA optimizes shared memory selection between PS and PL, while CHARM enhances PL-AIE communication efficiency. We add the BF16 support in CHARM for the following hardware-aware quantization (detailed in Section \ref{section_quant_algo}). For design space exploration (DSE) on PL, we configured TAPCA with multiple optimization directives. Table \ref{tab_design_points} specifies the selected design points and their configurations in AP-DRL. For loop unrolling, we sampled $\log_2LB$ points (where $LB$ represents loop bounds) in exponential progression across partitioning units to balance modeling accuracy and exploration efficiency. Regarding array partitioning, the maximum partitioning factor was constrained by the PS-PL-AIE memory interface maximum bitwidth $B_M$, calculated as $\lfloor B_M/B_D \rfloor$ where $B_D$ denotes the data bitwidth.

\subsection{Partitioning Algorithm}\label{section:partitioning_algo}

To fully exploit hardware parallelism for DRL training acceleration, we model the DRL training process as an Integer Linear Programming (ILP) formulation, facilitating automatic task partitioning. The DRL training process involving $N$ network layers can be represented as a Directed Acyclic Graph (DAG) $G=(V,E)$, where $V$ denotes the set of computational nodes (layers) and $E$ captures their dependencies. For each node $i \in V$, $\Gamma_i^-,\Gamma_i^+\subseteq V$ represent its predecessor and successor nodes, respectively, $t_{ij}$ indicates its execution time on component $j$, and $S_i$ denotes its start time. Each component $j$ has a total resource capacity $A_j$, and $V_j \subseteq V$ represents the set of nodes allocated to component $j$. The resource requirement of node $i$ when executed on component $j$ is denoted by $a_{ij}$. The binary decision variable $x_{ij}$ determines whether node $i$ is partitioned to component $j$. The objective of the partitioning algorithm is to minimize the total training time $T$, as formally expressed in Equation \eqref{equ_objective}.

\begin{align}
    &\min T.\label{equ_objective}\\
    \mathrm{s.t.}\ \ \ \ 
        & T=\max(S_i+x_{ij}t_{ij});\label{eq:T}\\
        & \sum_{j}x_{ij}=1,&&j\in\{\text{PL,AIE}\};\label{equ_decision_var}\\
        & S_{n}\geqslant x_{ij}t_{ij}+\sum_{\forall k\in\Gamma^-_i}x_{kj}t_{kj},&&n\in\Gamma_i^+;\label{equ_order}\\
        & T\geqslant S_i+x_{ij}t_{ij},&&\{i\in V\mid\Gamma_i^+=\varnothing\};\label{equ_make_sure}\\
        & \sum_{i\in V_j}a_{ij}\leqslant A_j.\label{eq:resource}
\end{align}

In the proposed partitioning model, the total execution time $T$ is determined by the completion time of the last node, as formalized in Equation~\ref{eq:T}. The assignment constraint in Equation~\ref{equ_decision_var} ensures each node is allocated to exactly one component. Dependency relationships between nodes are enforced through Equation~\ref{equ_order}, while Equation~\ref{equ_make_sure} guarantees that the total runtime encompasses all layer computations. Furthermore, resource constraints for node execution are specified in Equation~\ref{eq:resource}.

This ILP-based automatic partitioning model achieves optimal parallelization of the DRL training process by simultaneously determining the most efficient component assignment for each computational node, preserving all layer dependencies during parallel execution, and satisfying hardware resource constraints while minimizing total runtime. The formulation ensures that all aspects of the training process, including computational dependencies, resource limitations, and execution scheduling, are systematically coordinated to maximize parallel efficiency.

\subsection{Hardware-Aware Quantization Algorithm}\label{section_quant_algo}

In this section, we address the \textbf{third} challenge outlined in the design principles above. We begin this section by analyzing the impact of varying data precision characteristics on quantization. Building upon this analysis, we introduce a quantization algorithm that leverages these precision-dependent features to harness the heterogeneity of Versal ACAP platforms for efficient mixed-precision training implementations.

\subsubsection{Data Precision Analysis}\label{section_data_prec}

\begin{table}[]
    \centering
    \caption{Comparison of FP16, FP32 and BF16 in Mixed-Precision Training}
    \label{table_fp32}
    \begin{tabular}{c c c c}
    \toprule
         & \textbf{FP16} & \textbf{FP32} & \textbf{BF16} \\
         \midrule
         \makecell[c]{\textbf{Binary Format}\\(Sign, Exp, Frac)} & (1, 5, 10) & (1, 8, 23) & (1, 8, 7) \\
         \midrule
         \makecell[c]{\textbf{Exponent Range}\\(Base-2)} & [-14, 15] & [-126, 127] & [-126, 127]\\
         \midrule
         \makecell[c]{\textbf{Fraction Precision}\\(Decimal Digital)} & \makecell[c]{7 bits ($\sim$3)} & \makecell[c]{10 bits ($\sim$7)} & \makecell[c]{23 bits ($\sim$2)}\\
         \midrule
         \makecell[c]{\textbf{Memory Usage}\\(Bytes/Parameter)} & 2 & 4 & 2 \\
         \midrule
         \textbf{\makecell[c]{Master Weight\\Backup Required?}} & \makecell[c]{Yes} & N/A & No\\
         \midrule
         \textbf{\makecell[c]{Loss Scaling\\Required?}} & Yes & N/A & No \\
    \bottomrule
    \end{tabular}
\end{table}

As discussed in Section \ref{section_quant_bgd}, BF16, FP16, and FP32 each offer distinct advantages for different applications in mixed-precision training scenarios. The comparative analysis is presented in Table \ref{table_fp32}, which reveals that BF16's identical exponent range to FP32 provides inherent numerical stability, obviating the need for conventional stabilization techniques like master weight backup and dynamic loss scaling, thereby avoiding their associated computational overhead. 

In contrast, FP16's substantially narrower exponent range of [-14, 15] necessitates the above stability techniques to prevent both underflow in small gradient values and overflow in large activations. These precision-dependent characteristics become particularly critical when implementing heterogeneous mixed-precision training acceleration on Versal ACAP platforms, where optimal performance requires careful coordination between different numerical formats to maintain computational accuracy while maximizing hardware efficiency.

\subsubsection{Quantization Algorithm}

\begin{algorithm}[tbh]\label{algo_mpt}
    \newcommand\mycommfont[1]{\small\textcolor{gray}{\textit{#1}}}
    \SetCommentSty{mycommfont}
    \SetNoFillComment
    \LinesNumbered

    \caption{Hardware-Aware Quantization Algorithm}\label{quant_algo}
    \KwIn{\begin{itemize}
        \item Quantization delay $T_d$,
        \item DRL network $f_\theta$ with parameters $\theta$
    \end{itemize}}
    \KwOut{Trained DRL network $f_{\theta,\text{trained}}$}

    \For{timestep $t=1\cdots T$} {
        \If{current node is an AIE node} {
        \tcc{Execute on AIE (BF16)}
            Execute \textit{current node} in BF16.
        }
        \Else {
            \tcc{Execute on FPGA (FP16)}
            Convert BF16/FP32 to FP16;\\
            Duplicate BF16/FP32 weight as master weight to backup;\\
            \texttt{MPT(FP16,BF16/FP32)};\\
            Convert FP16 results to BF16/FP32;\\
            Update master weight with BF16/FP32.
        }
    }
\end{algorithm}

\begin{figure}
    \centering
    \includegraphics[width=7.8cm]{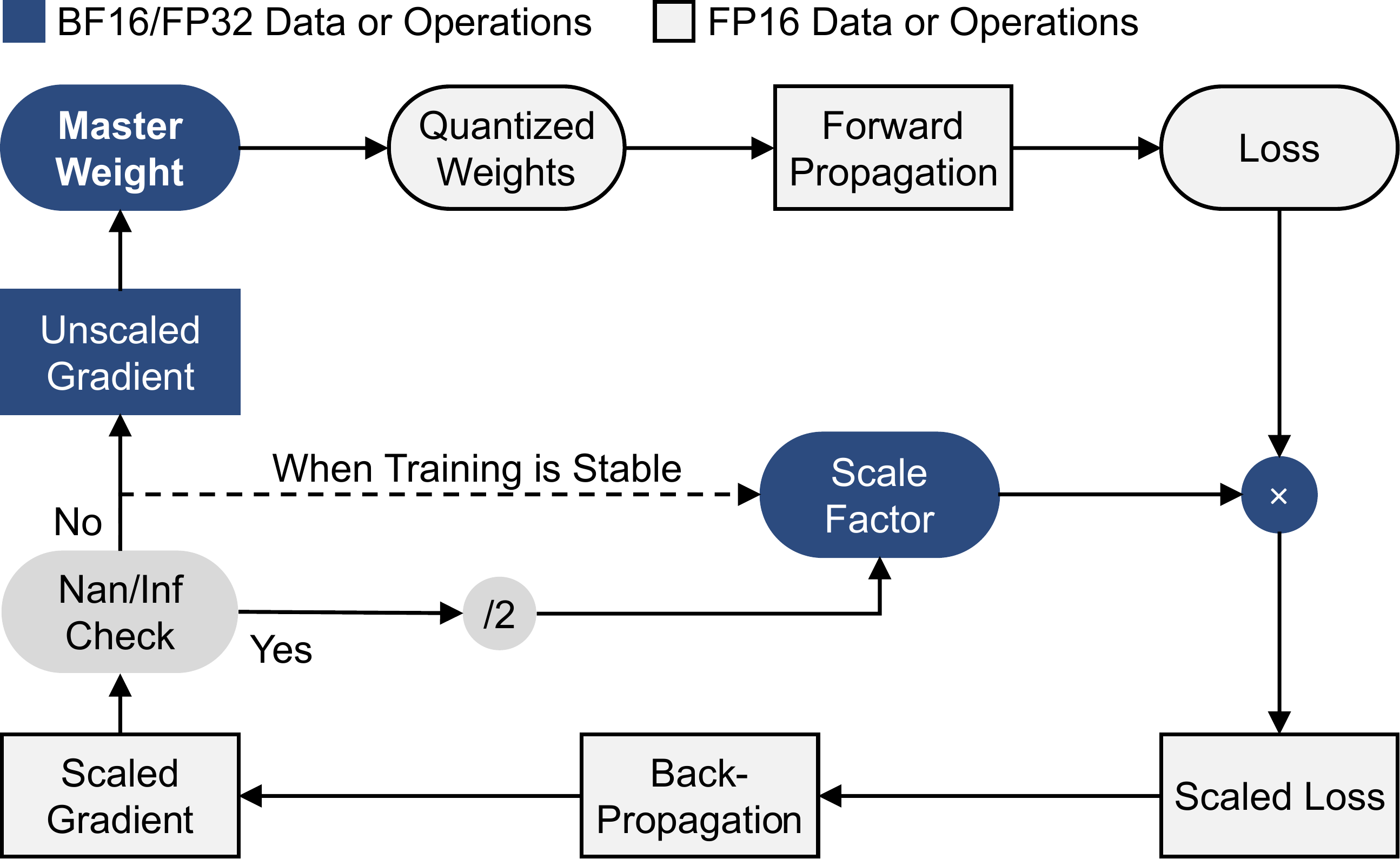}
    \caption{\texttt{MTP(FP16,BF16/FP32)} with the whole training steps in Algorithm \ref{algo_mpt}: Mixed-precision training workflow in FP16+BF16/FP32. Techniques of master weight backup and dynamic loss scaling are used.}
    \label{fig_mixed_prec_train}
\end{figure}

\begin{figure}
    \centering
    \includegraphics[width=6cm]{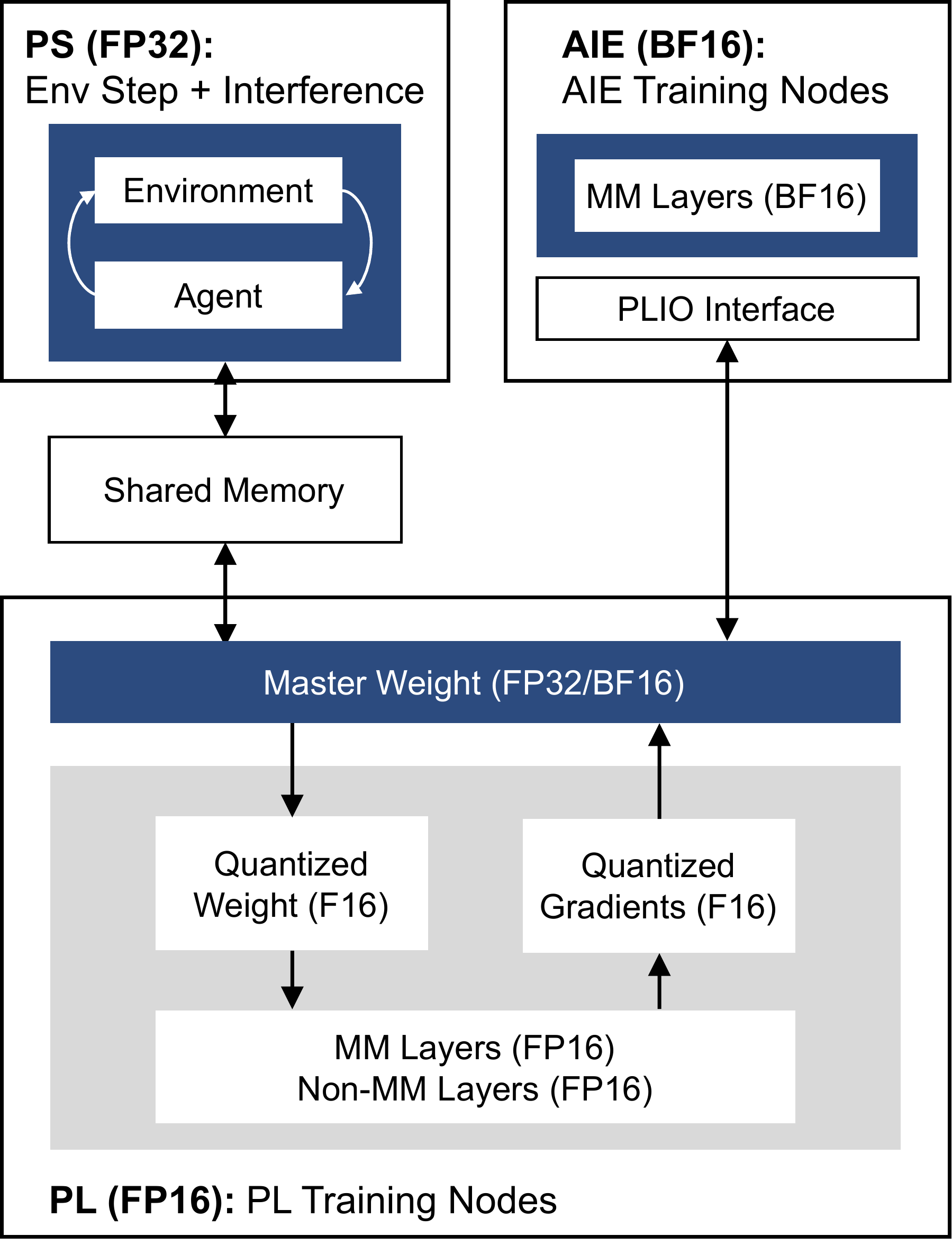}
    \caption{The hardware implementation of the quantization algorithm.}
    \label{fig:quant_hw}
\end{figure}

Based on the analysis of data precision characteristics presented in Section \ref{section_data_prec}, we propose a dynamic mixed-precision training algorithm specifically designed for the Versal ACAP heterogeneous computing platform, as described in Algorithm \ref{algo_mpt}. During DRL training, the algorithm dynamically adapts its precision strategy according to the hardware allocation of each computational node. For nodes mapped to AIE, all operations, including forward propagation, backpropagation, and weight updates, are performed in BF16 precision, leveraging its extended exponent range to avoid underflow without requiring additional optimization techniques. In contrast, nodes implemented on PL employ a carefully designed FP16+BF16/FP32 mixed-precision scheme that incorporates two critical optimizations: master weight backup maintained in higher precision (BF16/FP32) and dynamic loss scaling to prevent numerical underflow in FP16 computations.

The whole workflow of mixed-precision training with FP16+BF16/FP32 is presented in Figure \ref{fig_mixed_prec_train}: master weights are first converted to FP16 format, followed by forward propagation with dynamically scaled loss values to maintain proper numerical representation in FP16. The backward pass includes rigorous gradient validation (checking for NaN/Inf values) and conditional update skipping when overflow is detected, coupled with automatic scale factor adjustment. Valid gradients undergo precision conversion and proper unscaling before being applied to update the master weights.

This process is further optimized through hardware-aware optimizations tailored for Versal ACAP's heterogeneous architecture, as shown in Figure \ref{fig:quant_hw}. Using task partitioning, computational nodes allocated to the PL often handle only partial segments of the above process, requiring intelligent precision adaptation based on specific node functions. This adaptive approach employs distinct mixed-precision strategies for data compatibility: FP32+FP16 for nodes interfacing with PS, and BF16+FP16 for AIE interactions. The PL dataflow implementation incorporates synchronized master weight management, where FP32/BF16 master weight backups accompany input data streams, while output FP16 results undergo precision conversion to FP32/BF16 format before master weight updates. Special consideration is given to PL nodes handling loss function computations, where the integration of dynamic loss scaling is critical for maintaining numerical stability across the mixed-precision training pipeline.
\section{Experiments and Results}\label{section:experiments}

\subsection{Experiment Setup}

\begin{table*}[htb]
    \centering
    \caption{Experiment Configurations and Results}
    \label{table_benchmark_config}
    \begin{tabular}{llcccc}
    \toprule
        \makecell[c]{\textbf{Environment}} & \makecell[c]{\textbf{Algorithm}} & \makecell[c]{\textbf{Environment}\\\textbf{Space} $\left(\lvert\mathcal{S}\rvert,\lvert\mathcal{A}\rvert\right)$} & \makecell[c]{\textbf{Network Architecture}}  & \makecell[c]{\textbf{Train FLOPs}\\\textbf{(Per Batch Size)}} & \makecell[c]{\textbf{Reward Error}\\\textbf{(\%)}} \\
        \midrule
        CartPole & DQN & $(4,2)$ & \makecell[l]{3-layer MLP\\Hidden layer 1 size: 64\\Hidden layer 2 size: 64} & 28.04 K & 1.60 \\
        \midrule
        InvPendulum & A2C & $(4,1)$ & \makecell[l]{3-layer MLP\\Hidden layer 1 size: 64\\Hidden layer 2 size: 64} &  2.31 K & 4.81 \\
        \midrule
        LunarCont & DDPG & $(8,2)$ & \makecell[l]{3-layer MLP\\Hidden layer 1 size: 400\\Hidden layer 2 size: 300} & 2.25 M & 1.73 \\
        \midrule
        MntnCarCont & DDPG & $(2,1)$ & \makecell[l]{3-layer MLP\\Hidden layer 1 size: 400\\Hidden layer 2 size: 300} & 2.19 M & 1.12 \\
        \midrule
        Breakout & DQN & $(84\times84\times4,4)$ & \makecell[l]{ConvNet+MLP\\Conv1(4, 32, krnl\_size=8, stride=4)\\Conv2(32, 64, krnl\_size=4, stride=2)\\Conv3(64, 64, krnl\_size=3, stride=1)\\FC1(3136, 512), FC2(512, 4)} & 68.17 M & 1.25 \\
        \midrule
        MsPacman & PPO & $(84\times84\times4,9)$ & \makecell[l]{ConvNet+MLP\\Conv1(4, 32, krnl\_size=8, stride=4)\\Conv2(32, 64, krnl\_size=4, stride=2)\\Conv3(64, 64, krnl\_size=3, stride=1)\\FC1(3136, 512), FC2(512, 9)} & 106.23 M & 1.13 \\
    \bottomrule
    \end{tabular}
\end{table*}

We systematically evaluate AP-DRL through two key dimensions: (1) quantization effects and (2) task partitioning acceleration. The quantization analysis examines two critical metrics: average reward and training efficiency caused by quantization. Meanwhile, the task partitioning evaluation quantifies the actual speedup achieved by the AP-DRL framework.

To evaluate the effectiveness of AP-DRL, we conduct extensive experiments across three benchmark suites (Atari, OpenAI Gym, and MuJoCo), employing diverse environments:

\begin{itemize}
    \item \textbf{CartPole}: Classic control task in discrete action space.
    \item \textbf{InvertedPendulum (InvPendulum)}: Motor control task maintaining an unstable pendulum upright in continuous space.
    \item \textbf{LunarLanderContinuous (LunarCont)}: Rocket landing challenge requiring precise thruster control with multi-axis thrust adjustment.
    \item \textbf{MountainCarContinuous (MntnCarCont)}: Energy accumulation problem where a car must escape a valley in continuous space.
    \item \textbf{Breakout}: Arcade-style brick-breaking game testing reactive control in discrete action space with high FLOPs.
    \item \textbf{MsPacman}: Complex maze navigation with dynamic ghost avoidance in discrete action space with very high FLOPs.
\end{itemize}

 We evaluate four classic DRL algorithms (DQN, DDPG, A2C, and PPO) across all environments. The detailed experimental configurations are presented in Table \ref{table_benchmark_config}, where $\lvert\mathcal{S}\rvert$ denotes the dimension of the state space and $\lvert\mathcal{A}\rvert$ denotes the dimension of the action space. All evaluations employed hardware emulation on the VEK280 platform, which comprises a dual-core Cortex-A72 APU, 304 AIE-ML tiles, 1312 DSP engines, 520.7K LUTs, and 113.4 Mb PL memory.

\subsection{Quantization Effect}

\begin{figure}
    \centering
    \includegraphics[width=8.8cm]{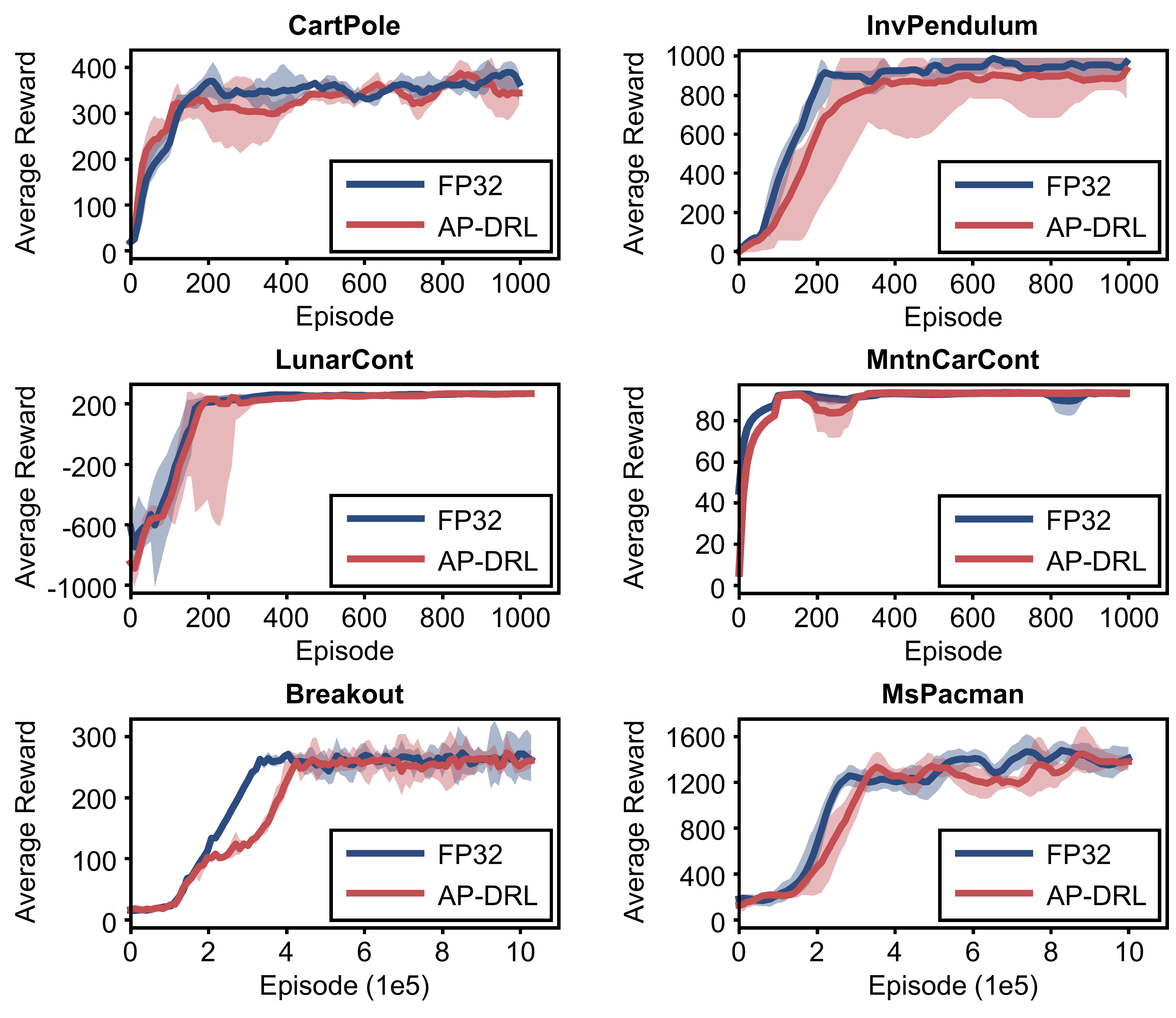}
    \caption{Convergence performance and average reward of AP-DRL. Solid curves represent the mean values across 5 independent runs with different random seeds, while shaded regions indicate the corresponding standard deviations.}
    \label{fig_quant_effect_convergence}
\end{figure}

To reflect the trend of reward changes clearly while minimizing the impact of data fluctuations, we calculate moving averages of episodic rewards using a 100-episode sliding window. To isolate the quantization effects, we conduct a controlled comparison between AP-DRL's quantized (mixed-precision with BF16/FP16/FP32) and non-quantized (FP32) implementations. Figure \ref{fig_quant_effect_convergence} illustrates the convergence behavior, while Table \ref{table_benchmark_config} quantifies the corresponding reward errors.

As illustrated in Figure \ref{fig_quant_effect_convergence} and Table \ref{table_benchmark_config}, AP-DRL successfully enables all models to converge to correct average rewards with high accuracy (reward error ranging from 1.12\% to 4.81\%). However, certain models exhibit slower convergence rates (notably the InvPendulum and DQN-Breakout), which may result from model-specific weight distributions, where wider distributions are more sensitive to the precision loss introduced by quantization.

To evaluate the impact of quantization on training efficiency, we conduct experiments with DQN-CartPole using network architectures of varying FLOPs as a case study. The results demonstrate that quantization delivers significant benefits as network FLOPs increase. This improvement stems from AP-DRL's tendency to allocate more layer nodes to the AIE in higher-FLOPs networks, where native BF16 support provides substantial performance gains. Conversely, networks with lower FLOPs predominantly distribute layer nodes to the PL. In low-FLOP scenarios, the computing time of processing units could fail to adequately overlap with master weight synchronization between PL and AIE, making this synchronization overhead non-negligible. As shown in Table \ref{table_quant_effect_overhead}, the master weight synchronization reduces BF16 quantization performance by at least 22\%.

\begin{table}[]
    \centering
    \caption{Training Time in One Episode for DQN-CartPole Using AP-DRL Across Network Architectures.}
    \label{table_quant_effect_overhead}
    \begin{tabular}{cccc}
    \toprule
        \textbf{\makecell[c]{Network\\Architecture\\(Hidden Size)}} & \textbf{\makecell[c]{Training Time\\(FP32)}} & \textbf{\makecell[c]{Training Time\\(BF16)}} & \textbf{Speedup} \\
        \midrule
         \makecell[c]{3-Layer MLP\\(64, 64)}& 0.21 ms & 0.27 ms & 0.78$\times$\\
         \midrule
         \makecell[c]{3-Layer MLP\\(400, 300)}& 32.55 ms & 28.81 ms & 1.13$\times$\\
         \midrule
         \makecell[c]{3-Layer MLP\\(4096, 3072)}& 2175.12 ms & 729.91 ms & 2.98$\times$\\
    \bottomrule
    \end{tabular}
\end{table}

\subsection{End-to-End Speedup}

\begin{figure}
    \centering
    \includegraphics[width=8.8cm]{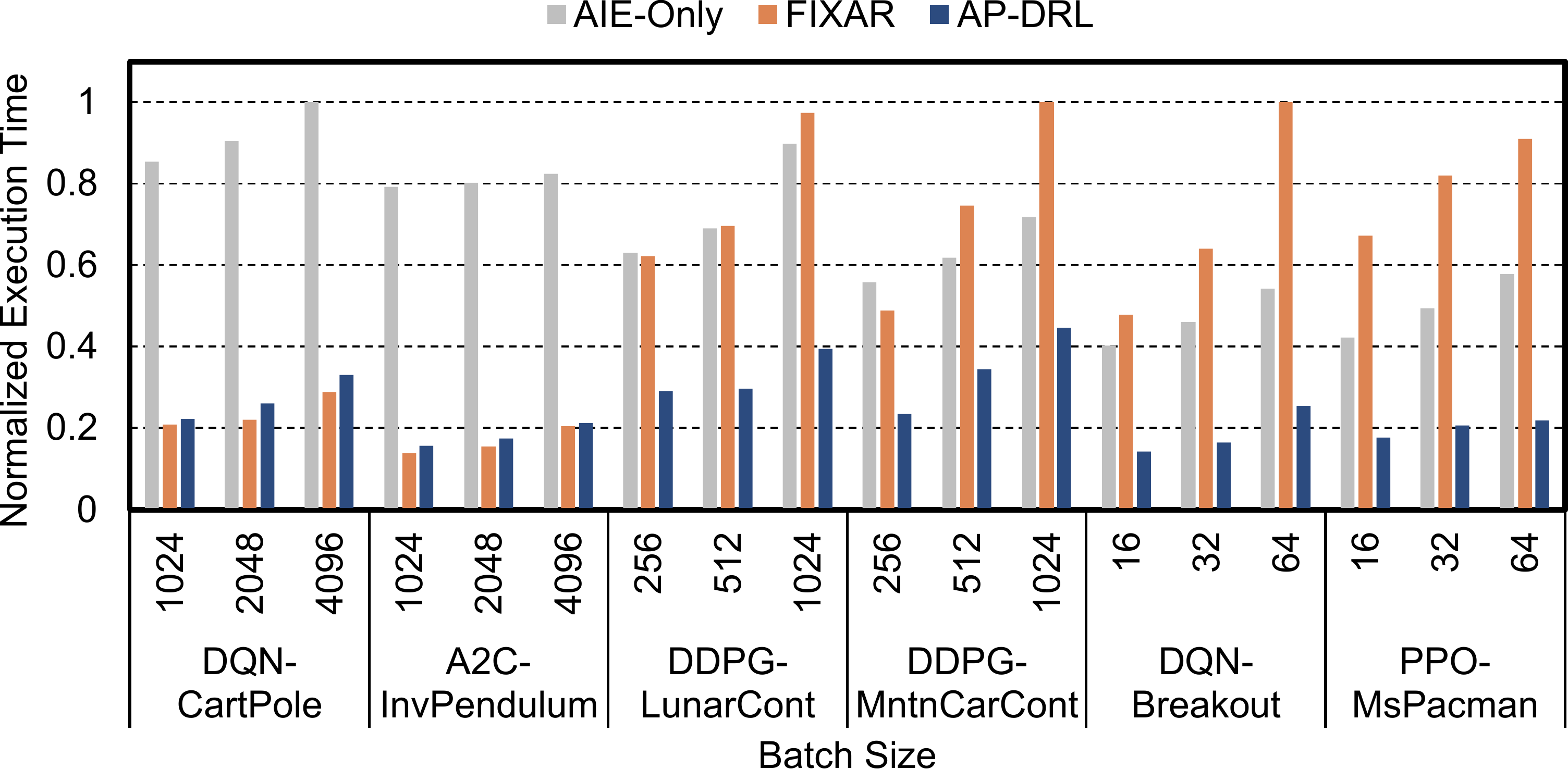}
    \caption{Normalized execution time of AIE-only, FIXAR, and AP-DRL across environments, algorithms, and batch sizes. Normalization is performed separately between DQN-CartPole and A2C-InvPendulum, DDPG-LunarCont and DDPG-MntnCarCont, and DQN-Breakout and PPO-MsPacman.}
    \label{fig:ap-drl_result}
\end{figure}

\begin{figure}
    \centering
    \includegraphics[width=8.8cm]{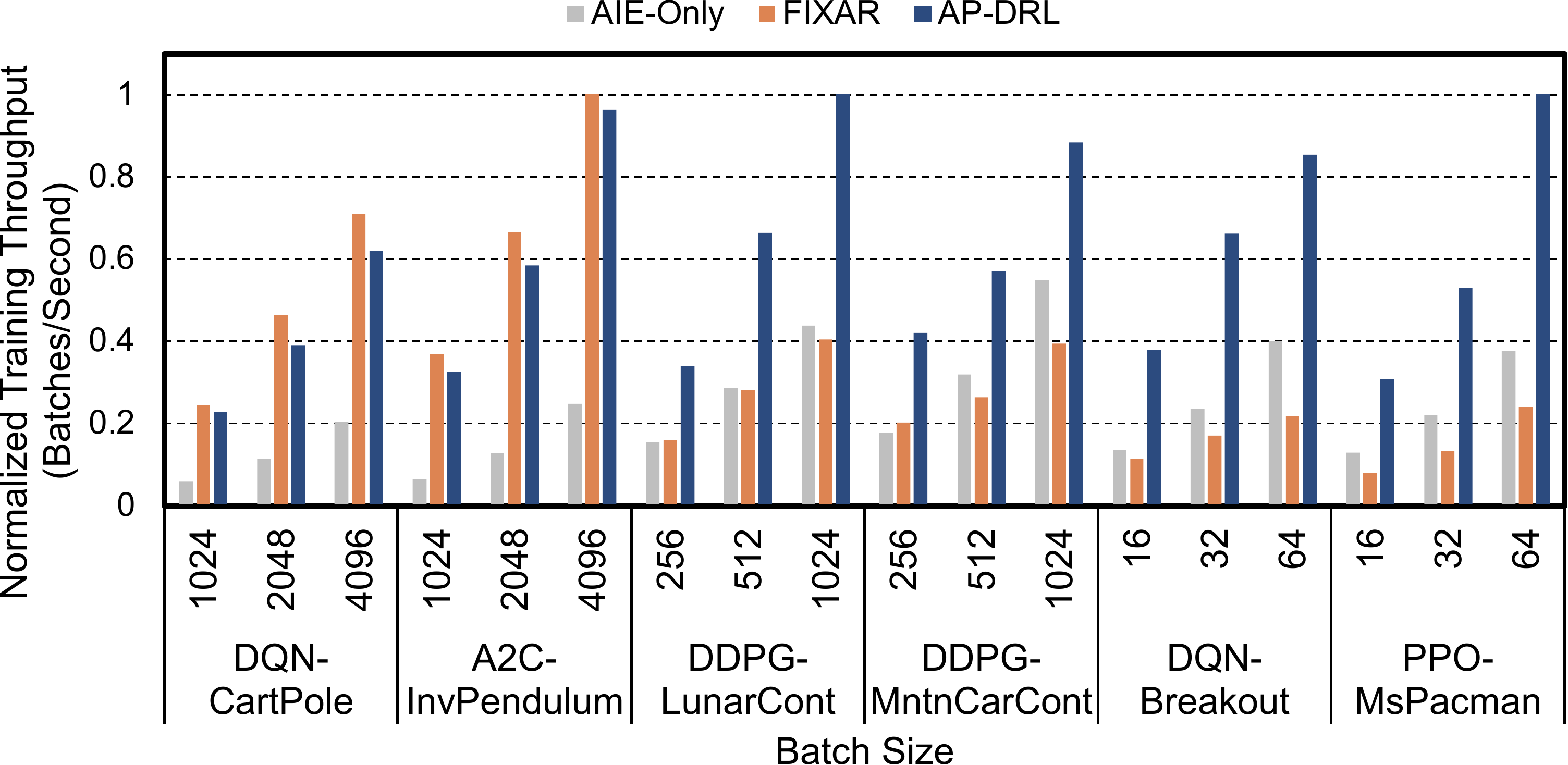}
    \caption{Normalized training throughput of AIE-only, FIXAR, and AP-DRL across environments, algorithms, and batch sizes. Normalization is performed separately between DQN-CartPole and A2C-InvPendulum, DDPG-LunarCont and DDPG-MntnCarCont, and DQN-Breakout and PPO-MsPacman.}
    \label{fig:ap-drl_result_throughput}
\end{figure}

 To demonstrate AP-DRL's acceleration efficiency, we compare it against two baselines: (1) an FP32 AIE-only scenario deployed using CHARM-optimized configurations, and (2) the FIXAR framework \cite{fixar}, the state-of-the-art heterogeneous acceleration methods for DRL training on the CPU-FPGA platform with fixed point quantization. For a comprehensive evaluation, each algorithm is tested with three distinct batch sizes, using \textbf{total training time} and \textbf{training throughput} as our primary performance metrics. Total training time refers to the sum of all step durations during the training phase within one timestep, while training throughput measures the number of training batches completed per second. Given the significant runtime variations (spanning orders of magnitude) across different environment-algorithm-batch size combinations, we apply normalization to enable meaningful cross-comparisons. All experiments are conducted with the PL and AIE operating at 245 MHz and 1 GHz, respectively.

 Since our experiments are conducted through hardware emulation, the power and resource utilization measurements may exhibit certain inaccuracies. Therefore, we do not present results related to power consumption and resource utilization.



\begin{figure}
    \centering
    \includegraphics[width=8.2cm]{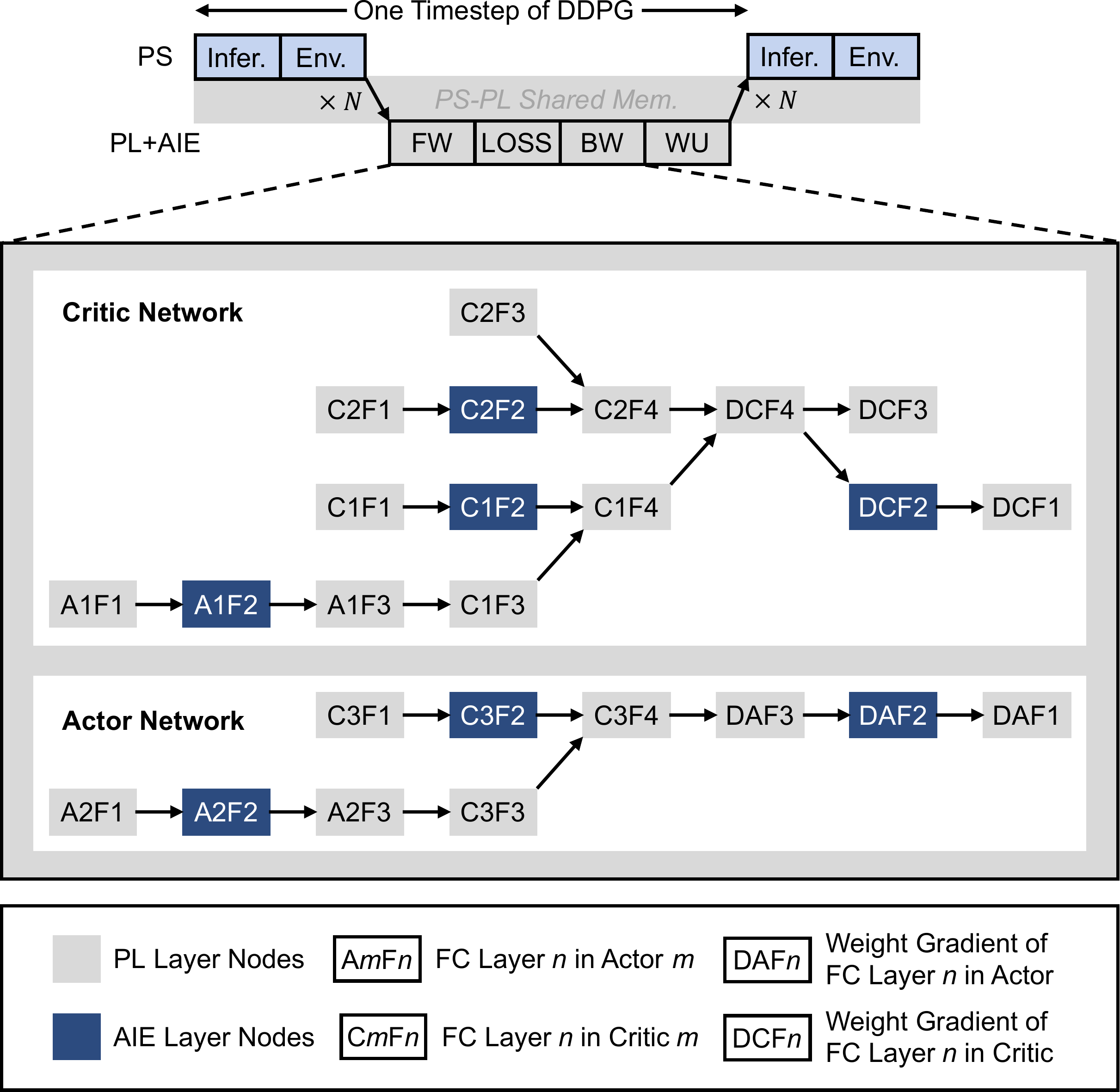}
    \caption{The operation sequence of DDPG-LunarCont in AP-DRL within one single timestep. The batch size of DDPG-LunarCont is 256.}
    \label{fig:case_study_ddpg256}
\end{figure}

\begin{figure}
    \centering
    \includegraphics[width=8.2cm]{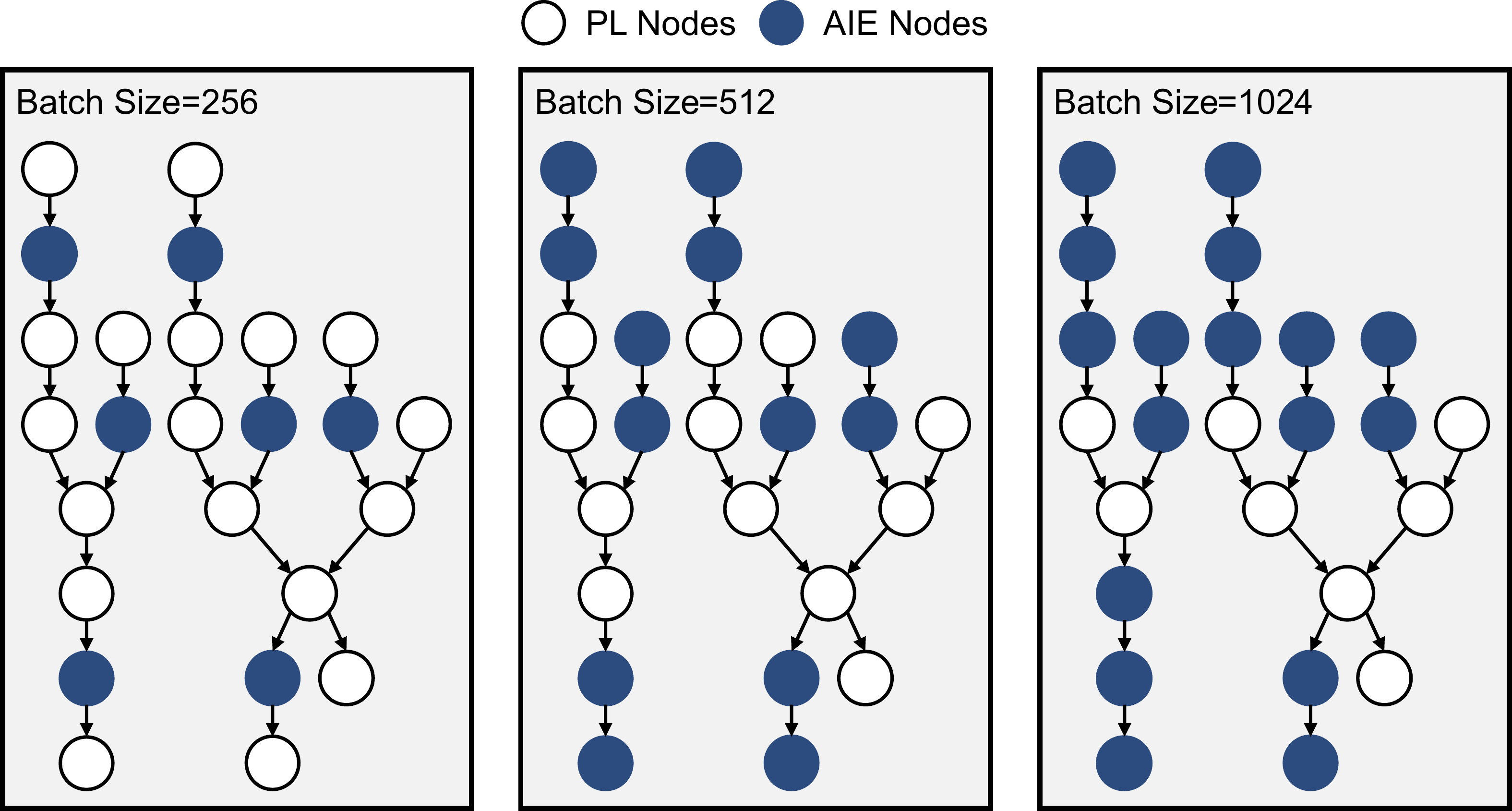}
    \caption{The partitioning results of AP-DRL on DDPG-LunarCont with batch sizes of 256, 512, and 1024. Each node represents a matrix multiplication layer (fully-connected layer in DDPG-LunarLander).}
    \label{fig:ap-drl_output}
\end{figure}

The experimental results are presented in Fig.~\ref{fig:ap-drl_result} and \ref{fig:ap-drl_result_throughput}. To further analyze the comparative performance between AP-DRL and the two baseline methods, we examine the DDPG-LunarCont case study, which demonstrates how AP-DRL's partitioning strategy evolves with increasing batch sizes and algorithm FLOPs. Specifically, we evaluate three batch size configurations (256, 512, and 1024), with results shown in Fig.~\ref{fig:case_study_ddpg256} and \ref{fig:ap-drl_output}. Our analysis reveals that as both batch size and algorithm FLOPs increase, AP-DRL progressively allocates more layer nodes to the AIE. This observation aligns with our findings in Section~\ref{section_bottleneck}: when the algorithm FLOPs is low, the significant initialization overhead of AIE results in inferior performance compared to the PL; however, at higher FLOPs, AIE's superior clock frequency enables better performance. Furthermore, the AIE's native hardware support for BF16 datatypes amplifies this performance advantage.

This analysis explains the results of performance comparison in Fig.~\ref{fig:ap-drl_result} and \ref{fig:ap-drl_result_throughput}:
\begin{itemize}
\item The AIE-only approach gradually outperforms FIXAR with increasing batch sizes and FLOPs, as the growing proportion of computation time reduces the impact of initialization overhead. When initialization becomes negligible, the AIE's higher clock frequency (1 GHz) delivers superior performance compared to FPGA-based FIXAR (164 MHz).
\item AP-DRL's performance advantage over FIXAR grows from marginal (0.98×) to substantial (4.17×) with increasing FLOPs. This trend arises because AP-DRL initially prioritizes PL allocation in low-FLOP regimes, where quantization synchronization bottlenecks constrain AP-DRL's performance, resulting in longer training time and lower training throughput compared to FIXAR. However, as FLOPs scales, AP-DRL transitions to AIE allocation, leveraging both clock frequency and quantization benefit to outperform FIXAR.
\item AP-DRL consistently outperforms AIE-only (1.61$\times$--3.82$\times$) across all FLOP ranges. This behavior occurs because at low FLOPs, AIE-only implementations suffer from significant initialization overhead, while AP-DRL leverages PL's shorter initialization time for better performance. As DRL FLOPs increase, AP-DRL allocates more layer nodes to AIEs. At this stage, AP-DRL outperforms AIE-only implementations due to its BF16 quantization optimization for layer nodes on AIEs.
\end{itemize}

\section{Conclusion}\label{section:conclusion}



Deep reinforcement learning (DRL) has found successful applications across numerous domains, while simultaneously presenting significant challenges in optimizing both quantization and automated task partitioning for training processes. In response, this paper introduces AP-DRL, an innovative framework that employs hardware-software co-design on the Versal ACAP platform to automate task partitioning and implement hardware-aware quantization optimization for DRL training.

Our methodology initiates with a comprehensive quantitative analysis of computing preferences across Versal ACAP's heterogeneous computing units under varying DRL configurations, establishing fundamental design principles. Based on these analytical insights, we develop a hardware-aware quantization algorithm based on mixed-precision training, subsequently integrating it with our proposed ILP-based partitioning model to create a unified optimization framework.

Through extensive experimentation across multiple DRL environments and algorithms, we validate AP-DRL's effectiveness with thorough comparisons against state-of-the-art approaches. The results indicate that the AP-DRL framework maintains training convergence while achieving performance improvements, with speedups reaching 4.17$\times$ compared to PL and 3.82$\times$ over AIE baselines. Future work will extend AP-DRL to incorporate DRL inference processes and environment steps, employing suitable quantization techniques to them to further enhance overall system performance.

\bibliographystyle{IEEEtranS}
\bibliography{refs}

\end{document}